%% file: main.tex
\documentclass[dvipsnames,twocolumn,onecolappendix]{aastex61}
\usepackage{aas_macros} 
\usepackage{natbib}
\usepackage{xspace}
\usepackage{fp}

\input{./measurements.tex}

\input{./calibration_constants.tex}
\input{./calculations.tex}
\input{./document_variables.tex}

\newcommand{\ho}{H\textsubscript{0}~}
\newcommand{\spliterr}[3]{ \( #1~\pm\) #2 \(\pm\) #3}
\newcommand{\hounits}{km s\(^{-1}\) Mpc\(^{-1}\)}
\newcommand{\LCDM}{\(\Lambda\)CDM~}

\shorttitle{TRGB distances to M66 and M96}
\shortauthors{Hoyt et al.}

\begin{document}

\title{The Carnegie Chicago Hubble Program VI. Tip of the Red Giant Branch Distances to M66 and M96 of the Leo I Group}

\author{Taylor J. Hoyt}
\affiliation{Department of Astronomy \& Astrophysics, University of Chicago, 5640 South Ellis Avenue, Chicago, IL 60637, USA}
\email{taylorjhoyt@gmail.com}

\author{Wendy L. Freedman}
\affiliation{Department of Astronomy \& Astrophysics, University of Chicago, 5640 South Ellis Avenue, Chicago, IL 60637, USA}

\author{Barry F. Madore}
\affiliation{The Observatories of the Carnegie Institution for Science, 813 Santa Barbara Street, Pasadena, CA 91101, USA}
\affiliation{Department of Astronomy \& Astrophysics, University of Chicago, 5640 South Ellis Avenue, Chicago, IL 60637, USA}

\author{Rachael L. Beaton}
\affiliation{Department of Astrophysical Sciences, Princeton University, 4 Ivy Lane, Princeton, NJ 08544, USA}
\altaffiliation{Carnegie-Princeton Fellow}
\altaffiliation{Hubble Fellow}
\affiliation{The Observatories of the Carnegie Institution for Science, 813 Santa Barbara Street, Pasadena, CA 91101, USA}

\author{Dylan Hatt}
\affiliation{Department of Astronomy \& Astrophysics, University of Chicago, 5640 South Ellis Avenue, Chicago, IL 60637, USA}

\author{In Sung Jang}
\affiliation{Leibniz-Institut f\"{u}r Astrophysik Potsdam, D-14482 Potsdam, Germany}

\author{Myung~Gyoon~Lee} \affiliation{Department of Physics \& Astronomy, Seoul National University, Gwanak-gu, Seoul 151-742, Republic of Korea}

\author{Andrew J. Monson}
\affiliation{Department of Astronomy \& Astrophysics, The Pennsylvania State University, 525 Davey Lab, University Park, PA 16802, USA}

\author{Jillian~R.~Neeley} \affiliation{Department of Physics, Florida Atlantic University, 777 Glades Road, Boca Raton, FL 33431, USA}

\author{Jeffrey A. Rich}
\affiliation{The Observatories of the Carnegie Institution for Science, 813 Santa Barbara Street, Pasadena, CA 91101, USA}

\author{Violet A. Mager}
\affiliation{Eberly College of Science, Penn State University Wilkes-Barre, Old Route 115, P.O. Box 264, Lehman, PA 18627, USA}

\begin{abstract}

We determine the distances to the Type Ia Supernova host galaxies M66 (NGC~3627) and M96 (NGC~3368) of the Leo~I Group using the Tip of the Red Giant Branch (TRGB) method. We target the stellar halos of these galaxies using the Hubble Space Telescope ACS/WFC in the F606W and F814W bandpasses. By pointing to the stellar halos we sample RGB stars predominantly of Population II, minimize host-galaxy reddening, and significantly reduce the effects of source crowding. Our absolute calibration of the $I$-band TRGB is based on a recent detached eclipsing binary distance to the Large Magellanic Cloud. With this geometric zero point in hand, we find for M66 and M96, respectively, true distance moduli \(\mu_0 =\)\spliterr{\MsixtysixDMRounded}{\MsixtysixCMBstaterrRounded (stat)}{\MsixtysixCMBsyserrRounded (sys)}~mag and \( \mu_0 = \)\spliterr{\MninetysixDMRounded}{\MninetysixCMBstaterrRounded (stat)}{\MninetysixCMBsyserrRounded (sys)}~mag.
\end{abstract}


\section{Introduction} \label{sect:intro}

There is an ever-increasing need for independent verification of the existing cosmic distance scale, rooted in the Cepheid Period-Luminosity relation, or the Leavitt Law. The tension between direct measurements and \LCDM inferences of the Hubble constant only appears to be growing \citep{fre17}. Locally determined estimates based on the Cepheid Leavitt Law currently find \ho\spliterr{= 73.0}{1.5 (stat)}{2.1 (sys)} \hounits ~ \citep{fre12} and \ho\(= 74.03~\pm \) 1.42 \hounits ~\citep{rie19}. By contrast, the \citet{pla18} find a \LCDM{} fit to the Cosmic Microwave Background (CMB) that yields an \ho{}\(= 67.8~\pm \) 0.9 \hounits. Independent of the CMB, \citet{des18} measure \ho \(= 67.2^{+1.2}_{-1.0} \) \hounits~via Baryon Acoustic Oscillations (BAO), paired with weak lensing and clustering data, and then anchored to independent Big Bang Nucleosynthesis (BBN) constraints. This puts the \LCDM inferred values of \ho in conflict with the direct, Cepheid-based measurements at the 3-4 sigma level. 

The precision of Cepheid-based measurements of the Hubble constant has slowly improved, but only at the tenth-of-a-percent level (e.g. 2.4\% to 2.1\% to 1.9\% precision) over the course of the last three years \citep{rie16,rie18, rie19}. Additionally, questions remain over the effect of metallicity on the zero point of the Cepheid Leavitt Law. For instance, see \citet{gie18} for a determination of a non-zero metallicity effect on the near-infrared (NIR) zero point. On the contrary, \citet{per04} saw no metallicity effect in the NIR.  Thus, it is critical that an independent distance scale validate, complement, and potentially improve upon the existing distance scale. To this end the Carnegie-Chicago Hubble Program (CCHP) is undertaking an independent calibration of the cosmic distance scale using a Population~II application of the Tip of the Red Giant Branch (TRGB), out to distances comparable to those already probed by the Population~I Cepheids. 

The TRGB is defined as the sharp decline in stellar source counts seen immediately at and above the peak luminosities of first-ascent red giant stars. Inside the RGB star, this sharp decrement is the result of a core helium flash, the seconds-long thermonuclear runaway resulting from the ignition of helium burning in the degenerate cores of low-mass red giant stars. For \(M \lesssim 1.4 M_{\odot} \) the bolometric luminosity of stars at the TRGB (i.e., at the time of the helium flash) is negligibly dependent on initial stellar mass, stellar age, and metallicity \citep{sal05}. Fortuitously, in the \(I\)-band the luminosity of stars at the TRGB is weakly dependent on metallicity and age, and by proxy, photometric color. However, at higher masses/younger ages the luminosity at the TRGB trends more strongly with parameters like metallicity and age. As such, mixed stellar populations (in particular the presence of contaminating young and intermediate-aged populations) would catastrophically preclude any application of the method to determine an accurate distance.

The CCHP was designed to expressly address all of these concerns and push the limits of the TRGB method by using the Hubble Space Telescope (HST) to resolve the stellar halos of galaxies that have been host to Type Ia supernovae (SNe~Ia) \citep{bea16}. By pointing to the stellar halos, where the \(M \lesssim 1.4 M_{\odot} \) restriction is fulfilled, we can expect a mostly homogeneous population of stars, on top of which these stars are also expected to be metal-poor, of Population~II. The CCHP has demonstrated the overall consistency and power of this halo-targeted application of the TRGB method by measuring distances precise to a few percent to galaxies both within 1~Mpc \citep[IC1613]{hat17} and out to as far as 20~Mpc \citep[NGC 1365]{jan18}. 

In this paper we present halo TRGB distances to M66 (NGC~3627) and M96 (NGC~3368). These are both spiral galaxies located in the Leo Constellation, and they are thought to be bound to the larger Leo~I Group \citep{dev75}. The pair of galaxies is relatively nearby, \( \sim 10 \)~Mpc, and have each recently hosted a SN~Ia event for which modern, CCD-based light curves are available.

Previously published as part of the CCHP are distances to NGC~1365 \citep{jan18}, NGC~4424, NGC~4526 and NGC~4536 in the Virgo cluster \citep{hat18a}, and NGC~1448, NGC~1316 of the Fornax cluster \citep{hat18b}. The distance to M101, the most nearby galaxy in the program, will be discussed in Beaton et al. (2019, submitted).

M66 and M96 were host to SN1989B and SN1998e, respectively, and were not included in the \cite{rie16} distance ladder because these SNe Ia were deemed to be too highly reddened (\(A_V > 0.5\)). However, by leveraging information contained in the multi-wavelength, optical to NIR, SN Ia light curves, one can explicitly solve for the SN internal extinction \citep[e.g.][]{bur11, bur14}, thus making these SNe~Ia viable in a precision determination of the Hubble constant. Both M66 and M96 are also the galaxies with the second and third highest metal content in the HST Key Project sample, requiring \(\gtrsim 0.2\) mag metallicity corrections to Cepheid-based distance measurements which were zero-point calibrated via the relatively metal-poor Large Magellanic Cloud (LMC). Lastly, Cepheid distances to this pair of Leo I group galaxies have proven to be highly dependent on the choice of period-selection cuts made before applying the Leavitt Law \citep{fre01,sah06}, leading to ambiguity in their published distances.

In M66 and M96, the dust content, high mean metallicity, and apparent dependence of Cepheid distance determinations on period cuts, makes them prime laboratories for demonstrating the consistency and accuracy of the TRGB. By targeting the stellar halos we simultaneously minimize both dust and metallicity effects, while the intrinsic nature of the TRGB necessarily avoids any periodicity-based selection.

\section{Data and Photometry} \label{sect:data}

\subsection{Observations} \label{subsect:obs}
\begin{figure*}
\includegraphics[width=\textwidth, trim={1cm 3cm 1cm 3cm},clip]{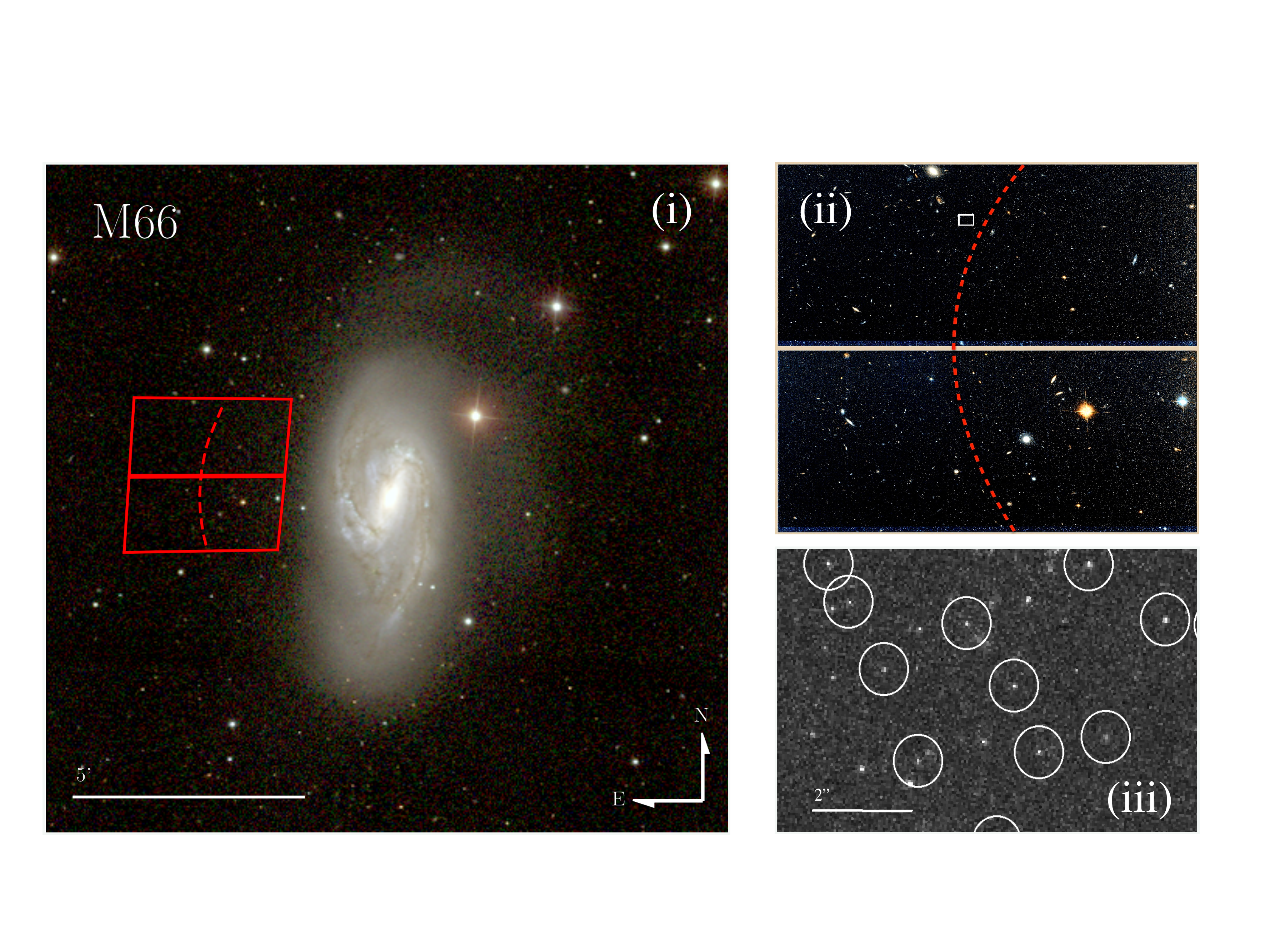} \\
\includegraphics[width=\textwidth, trim={1cm 3cm 1cm 3cm},clip]{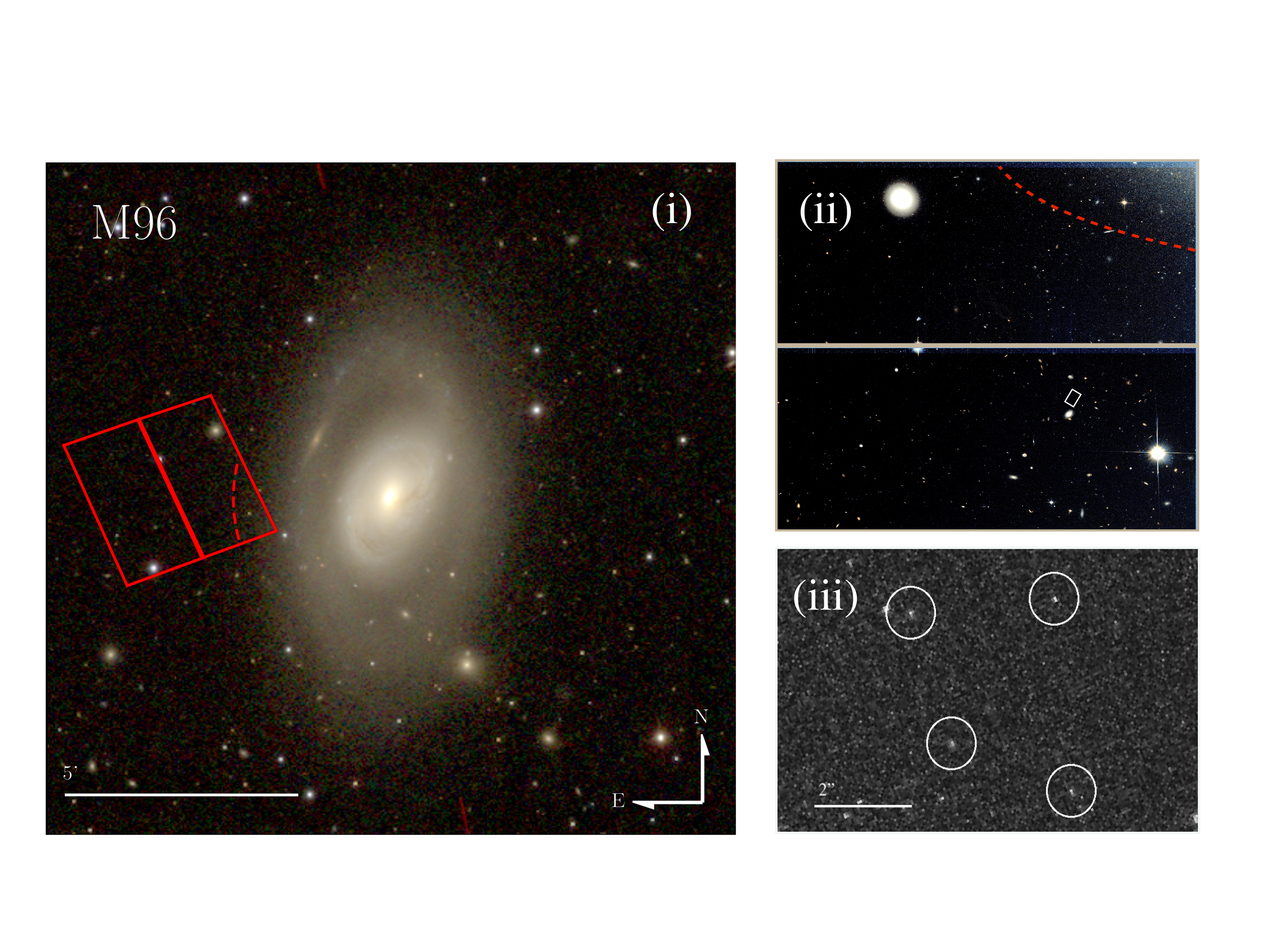}
\caption{(i) SDSS \textit{gri} cutouts for M66 and M96. In red are the HST ACS pointings used in the present study. The dashed red lines are lines of constant galactic radius within which all detected sources are excluded. These spatial cuts are discussed in detail in \autoref{subsect:spatialcuts}. (ii) Median color images of the ACS F606W and F814W pointings. (iii) 150x150pix cutouts of the median stack image. RGB stars down to 0.5 mag fainter than the detected TRGB are circled.}
\label{fig:pointing}
\end{figure*}

Observations were made with the ACS/WFC instrument on board the Hubble Space Telescope (HST) as part of a Cycle 22 program \cite[PID:GO13691]{ref:cchpprop}. In \autoref{tab:obs} we provide dates observed, exposure times taken in the F606W and F814W bandpasses, and the ACS pointing positions. In \autoref{fig:pointing} the ACS pointing footprints are outlined in solid red. The pointing positions were chosen per CCHP guidelines, described in \cite{bea16}, most importantly at radial distances that extend sufficiently out into the galaxy halo, away from spiral arms and UV-bright star forming regions, which minimizes both dust extinction and contamination by younger stellar populations.

M66 was observed on 10 March 2015, with two orbits dedicated to F814W imaging and one orbit for F606W. M96 was observed on 29 November 2015, with three orbits dedicated to F814W imaging and 1 orbit for F606W. We extracted from MAST the processed \textit{.flc} data products, which are flat-fielded and corrected for charge transfer inefficiencies (CTE-corrected). The \textit{.flc} images were furthered corrected by multiplying by their respective Pixel Area Maps\footnote{http://www.stsci.edu/hst/acs/analysis/PAMS} to account for the sky-projected pixel areas across the chips varying due to geometric distortions.

\input{./obs_log.tex}

\subsection{Photometry} \label{subsect:phot}

Our photometry proceeds similarly to previous CCHP papers \citep{hat17,jan18,hat18a}. In summary, we used theoretical PSFs as calculated with the \textit{TinyTim} PSF modeling package \citep{kri11}. We then used the TinyTim PSFs to photometer our images with the \textsc{DAOPHOT+ALLFRAME} suite \citep{ste87}, and the photometry was calibrated to the Vegamag system following standard procedures outlined by STScI. 

\begin{figure*}
\includegraphics[width = 0.49\textwidth]{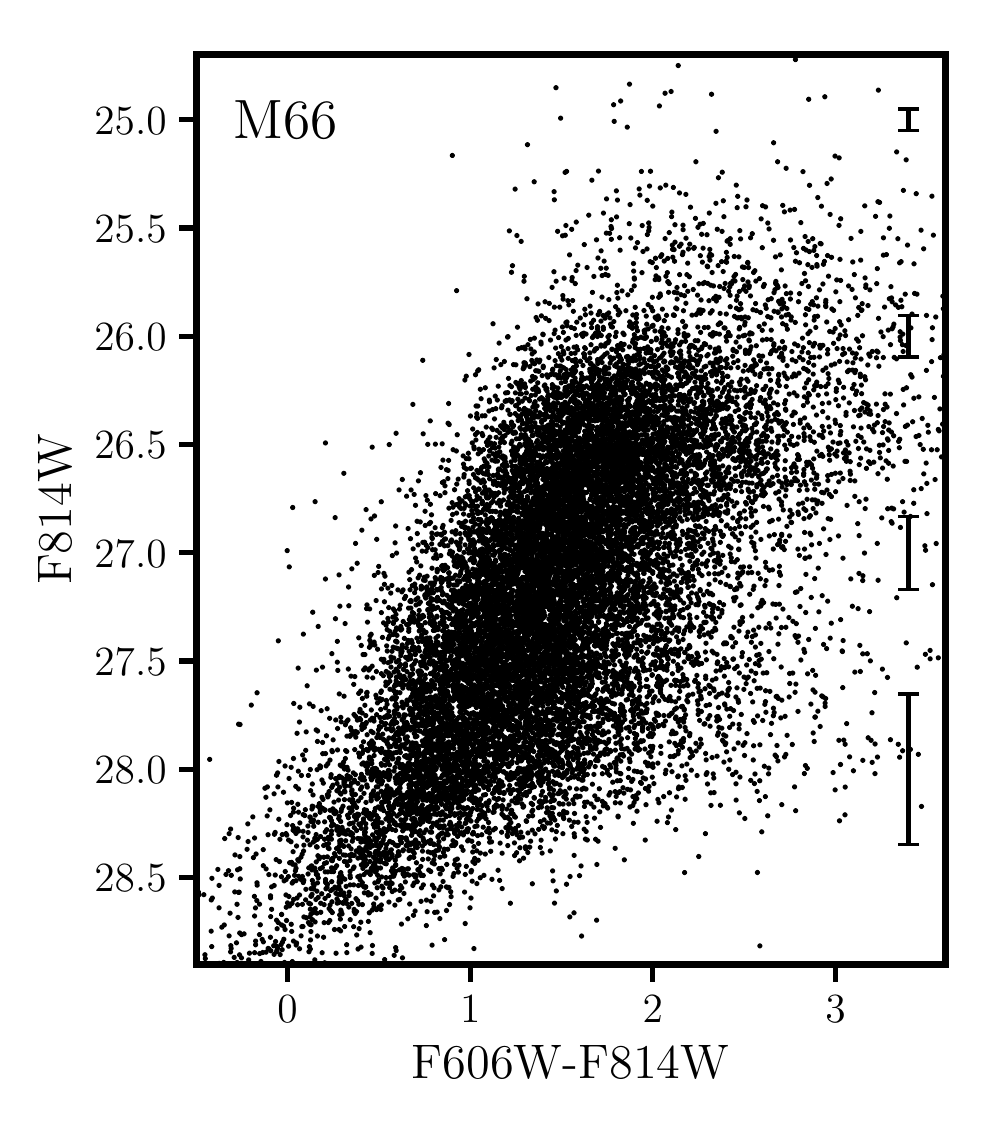}
\includegraphics[width = 0.49\textwidth]{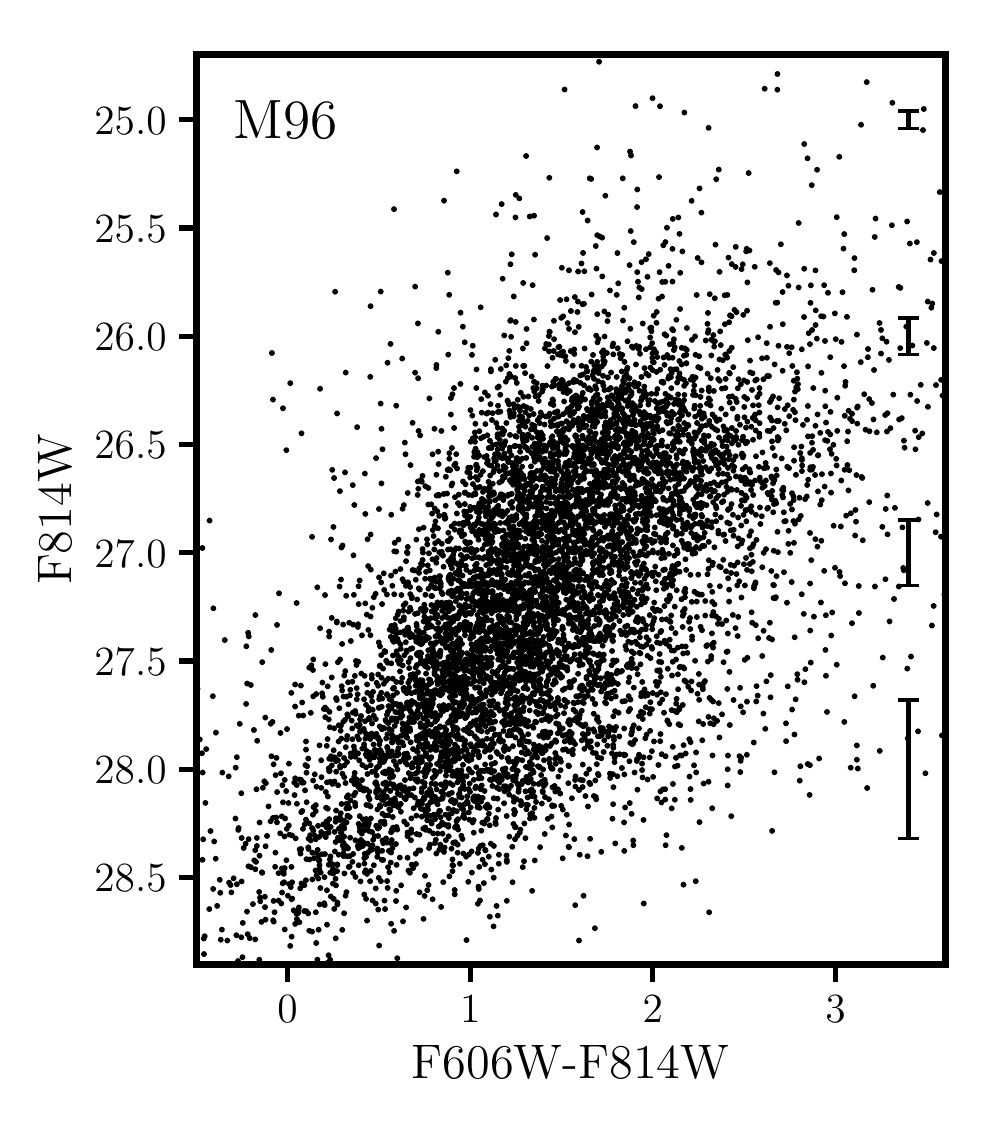}
\caption{CMDs for M66 (left) and M96 (right). Included are representative measurement uncertainties at regular 1 mag intervals. The sharp discontinuity in stellar counts that defines the TRGB is visible to the eye, somewhat fainter than F814W\(\sim 26\)~mag.
}
\label{fig:cmds}
\end{figure*}

This study makes use of an automated photometry pipeline outlined in Beaton et al. (2019, submitted) which follows the same principles as earlier papers in the CCHP series. Measured aperture corrections were applied on a frame-by-frame basis, and we present their average values in \autoref{tab:apcors}. Stellar source catalogs were then derived by applying automatically computed selection-cuts to the \textsc{daophot} morphological parameters \textsc{chi} and \textsc{sharp} in addition to a selection cut on signal-to-noise. The calibrated color-magnitude diagrams (CMDs) are given in \autoref{fig:cmds}.

\input{./apcors.tex}

\subsection{Artificial Star Tests} \label{subsect:artstars}

We generated artificial stars to assess our photometric accuracy for stellar sources in the magnitude range near the anticipated TRGB magnitude. We further used the artificial stars to construct synthetic CMDs as a means of quantifying both the statistical and systematic errors in our TRGB measurement. 

The artificial stars are sampled from a theoretical luminosity function (LF) that is comprised of an asymptotic giant branch (AGB) and a red giant branch (RGB) component. The slopes of the AGB and RGB LFs were assumed to be 0.1 and 0.3~dex/mag, respectively. The number of AGB to RGB stars at the anticipated TRGB magnitude is assumed to be 1:4, which is the mid-range abundance for observed populations in local galaxies
\citep[see, e.g.][]{ros14}. To begin the artificial star tests, we sampled from the theoretical LF 2000 stars, placed them uniformly into each image, and performed the same photometry routines as used on the unperturbed images. This process was continued for 250 iterations for each CCD, ultimately processing 1 million artificial stars for each galaxy. The purpose of the artificial star test is twofold: (1) To inform how we would expect the morphology of the TRGB to manifest in our real data; and (2) to estimate the uncertainties associated with measuring the TRGB in our dataset. In \autoref{subsect:errors} we discuss in detail the results of the artificial star tests.

\section{The TRGB distances to M66 and M96} \label{sect:trgb}

The TRGB is a precision distance indicator that traces its origins back to \cite{sha18}, who assumed that the brightest stars found in distant galactic globular clusters were of the same luminosity. \citet{baa44} later used the magnitude at which this so-called ``red sheet'' of Population~II stars first resolved out, to estimate distances to a number of Local Group galaxies. Subsequently, a number of TRGB distances were simply estimated by eye until \citet{lee93} established the modern $I$-band approach that involves using a first derivative, finite difference kernel to algorithmically measure the luminosity of the TRGB. Now the CCHP has adopted a generalized methodology that we describe briefly below \citep[see][for a more detailed discussion]{hat17}. In this section we first discuss the edge detection process, then the results of the artificial star tests used to evaluate error budget in our measurement of the TRGB, followed by discussion of the adopted spatially dependent masks, then ultimately we present the distances to M66 and M96.

\subsection{Detecting the Tip} \label{subsect:tip}
\begin{figure*}
\includegraphics[width = 0.49\textwidth]{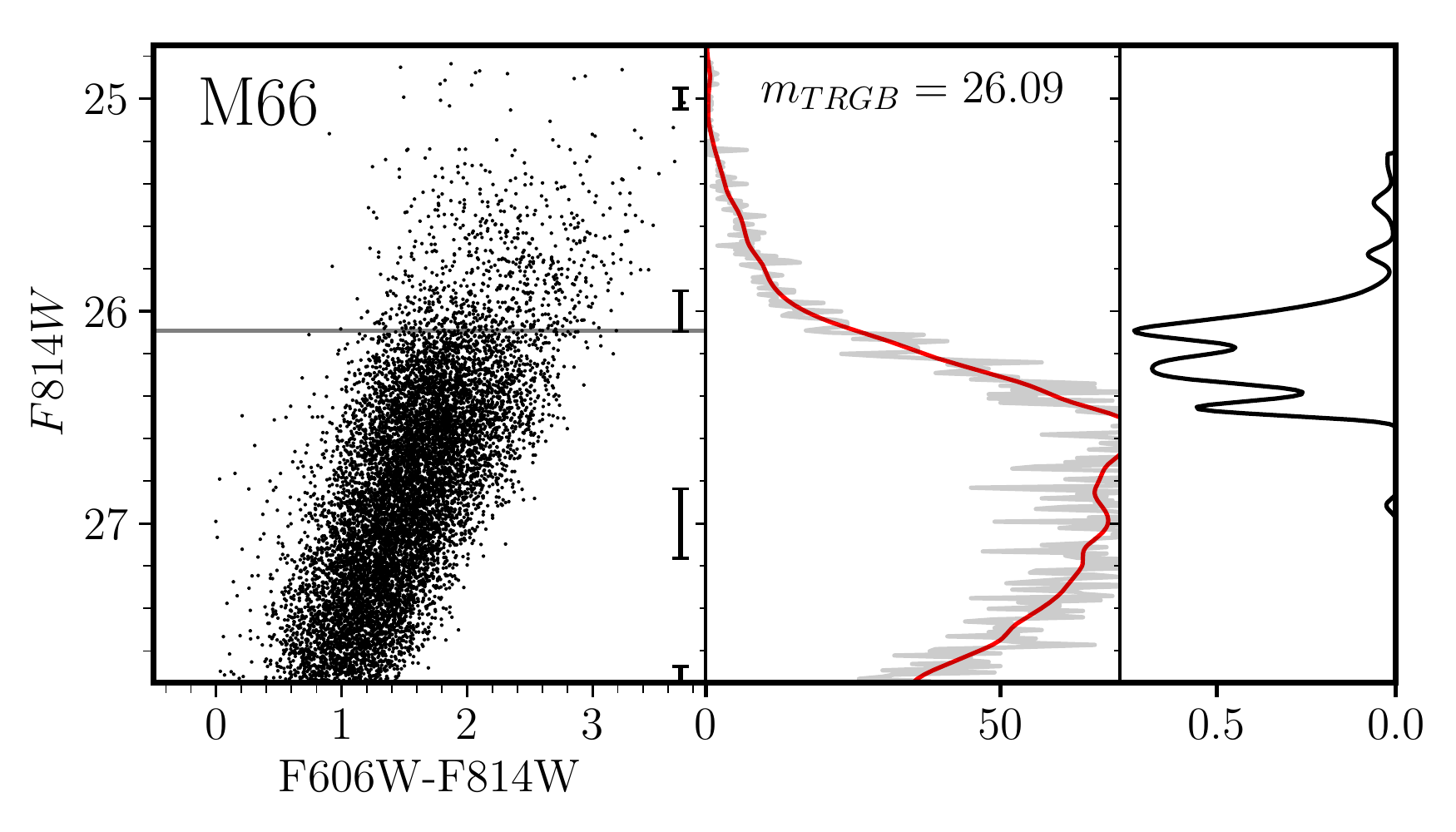}
\includegraphics[width = 0.49\textwidth]{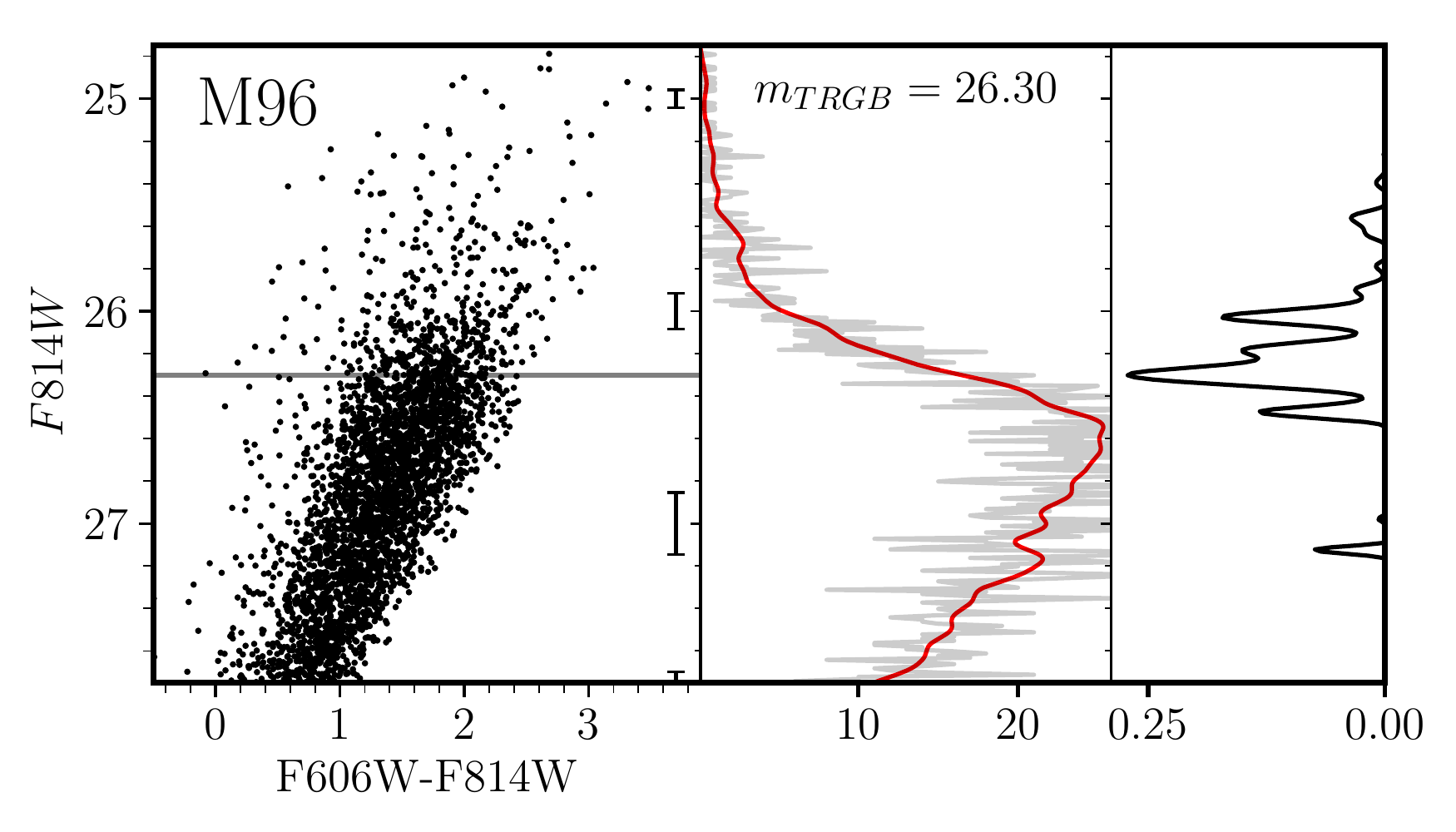}\\
\includegraphics[width = 0.49\textwidth]{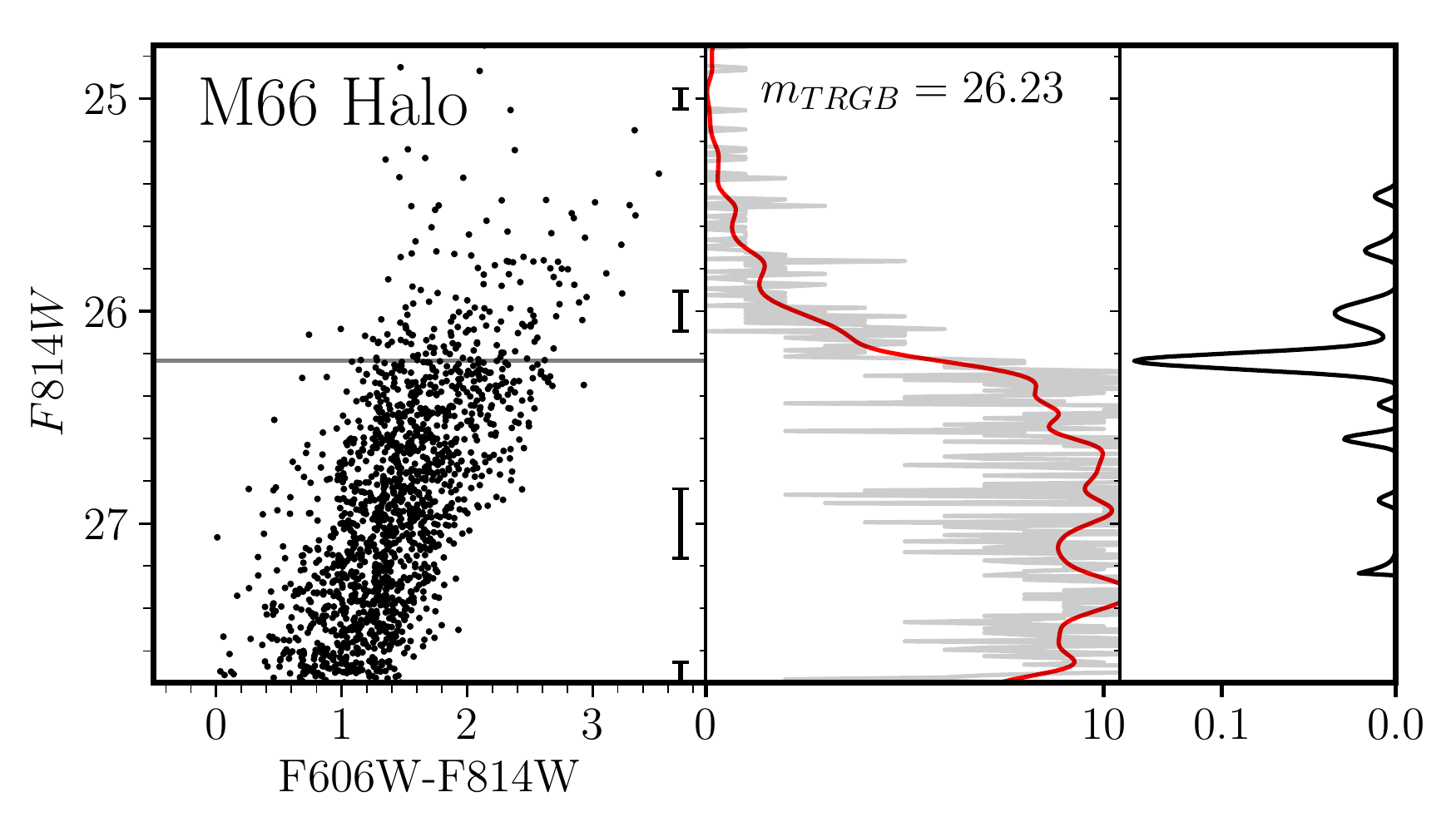}
\includegraphics[width = 0.49\textwidth]{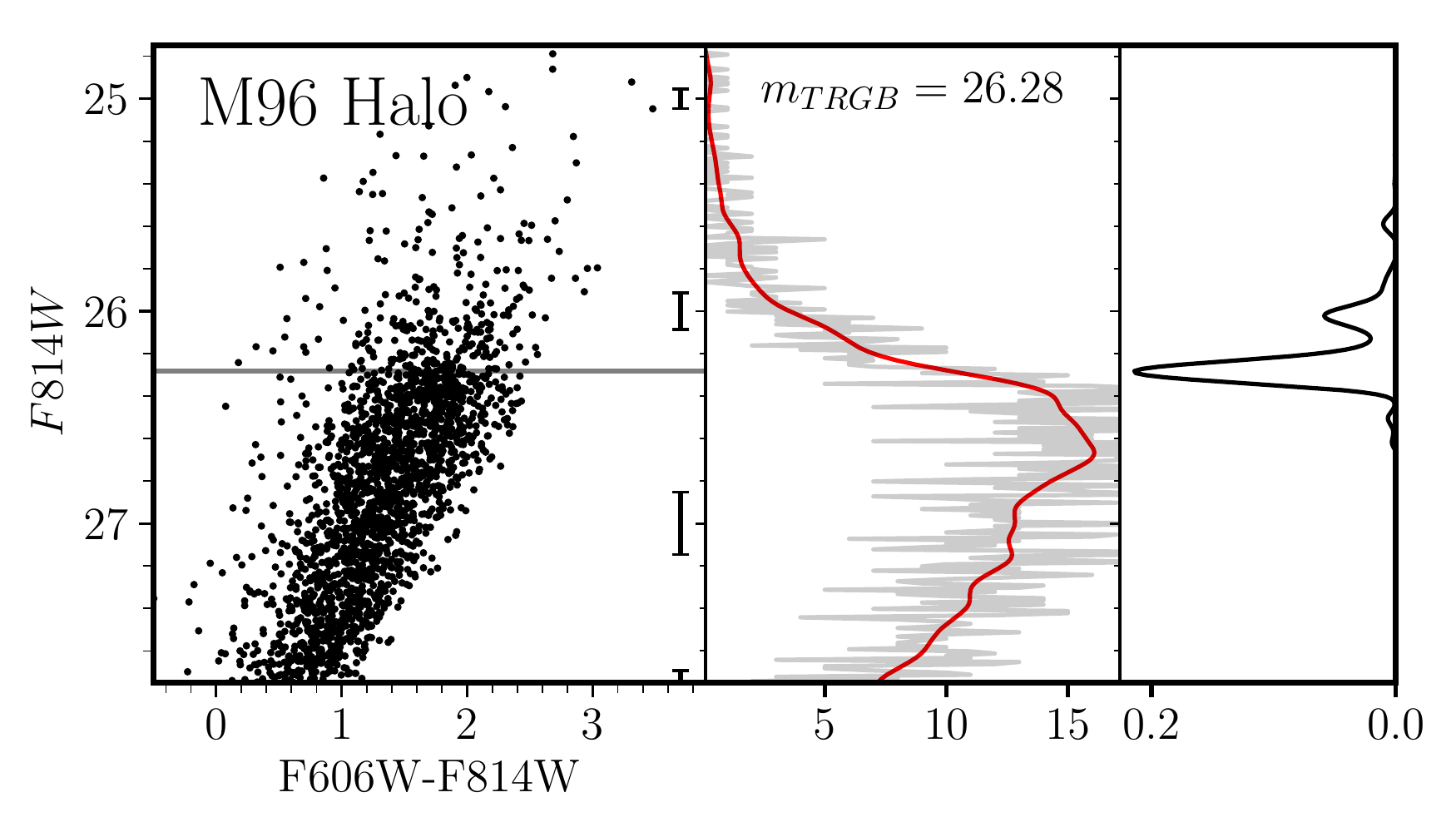}
\caption{Measurement of the TRGB in M66 and M96. Left column: plotted are the CMDs with a horizontal line at the detected TRGB magnitude. Middle column: the GLOESS (see text) smoothed luminosity function used for the edge detection is plotted in red. The original 0.01 mag binned LF is plotted underneath in gray. Right column: the Sobel filter $[-1, 0, +1]$ Poisson noise-weighted response function, the peak of which is taken to be at the detected magnitude of the TRGB.}
\label{fig:detection}
\end{figure*}

To measure the TRGB, we marginalize over color in an \(I\)-band CMD (left panels of \autoref{fig:detection}) to produce an $I$-band luminosity function (middle panels of \autoref{fig:detection}). The color baseline conventionally used, and that which we use here, to isolate the RGB, is \( ( V - I ) \). The luminosity function (LF) is first binned at the 0.01 mag level, and then smoothed using a non-truncated Gaussian kernel, called GLOESS \citep[see, e.g.,][]{per04,mon17}. Next, a Sobel $[-1, 0, +1]$ edge detection filter is passed over the smoothed LF to produce a trace of the first derivative of the luminosity function. To produce the final response function (right panels of \autoref{fig:detection}), we weight the first derivative by the Poisson noise of adjacent bins in the smoothed LF.\footnote{This weighting scheme is ideal for suppressing purely statistical fluctuations in sparsely populated regions of the LF, specifically above the TRGB where the number of AGB stars drops precipitously relative to the number statistics of the RGB.} The magnitude at which this signal-to-noise weighted, first-derivative response function reaches its maximum marks the TRGB.

\subsection{TRGB Measurement Error} \label{subsect:errors}
Because the entire RGB luminosity function contributes to a single measurement of the TRGB, it is not immediately obvious how to evaluate the errors associated with our edge detection methodology. The width of the TRGB peak in the response function will surely be an overestimate, being that it is a function of both the bin sizes and smoothing kernel width chosen. One could also consider the width of the peak response divided by the square root of the number of stars in the corresponding bins in the LF, as a proxy for an error on the mean. However, this definition is too open-ended, relying heavily on the user to choose a ``reasonable'' smoothing width.

\begin{figure*}
\includegraphics[width = 0.49\textwidth]{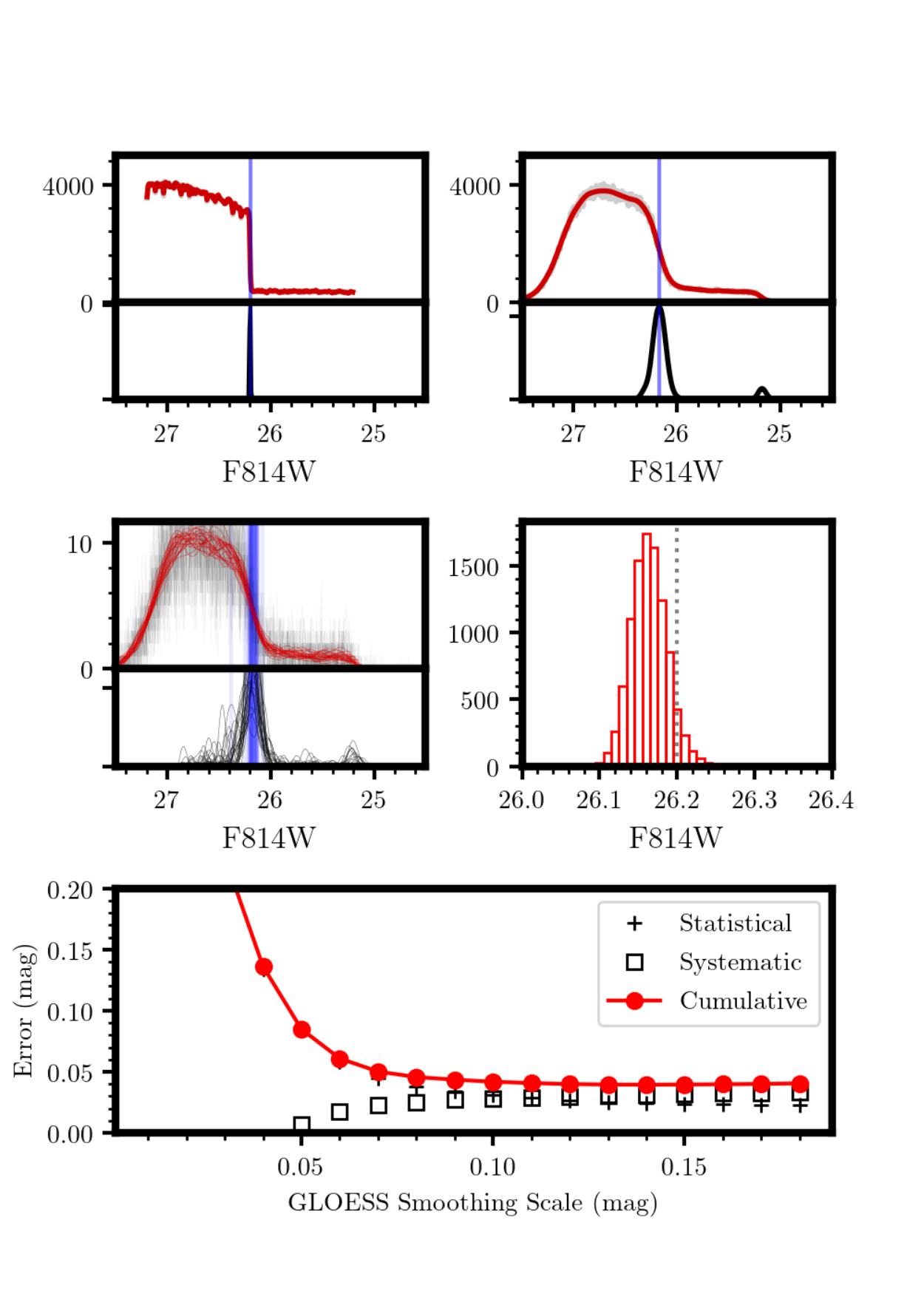}
\includegraphics[width = 0.49\textwidth]{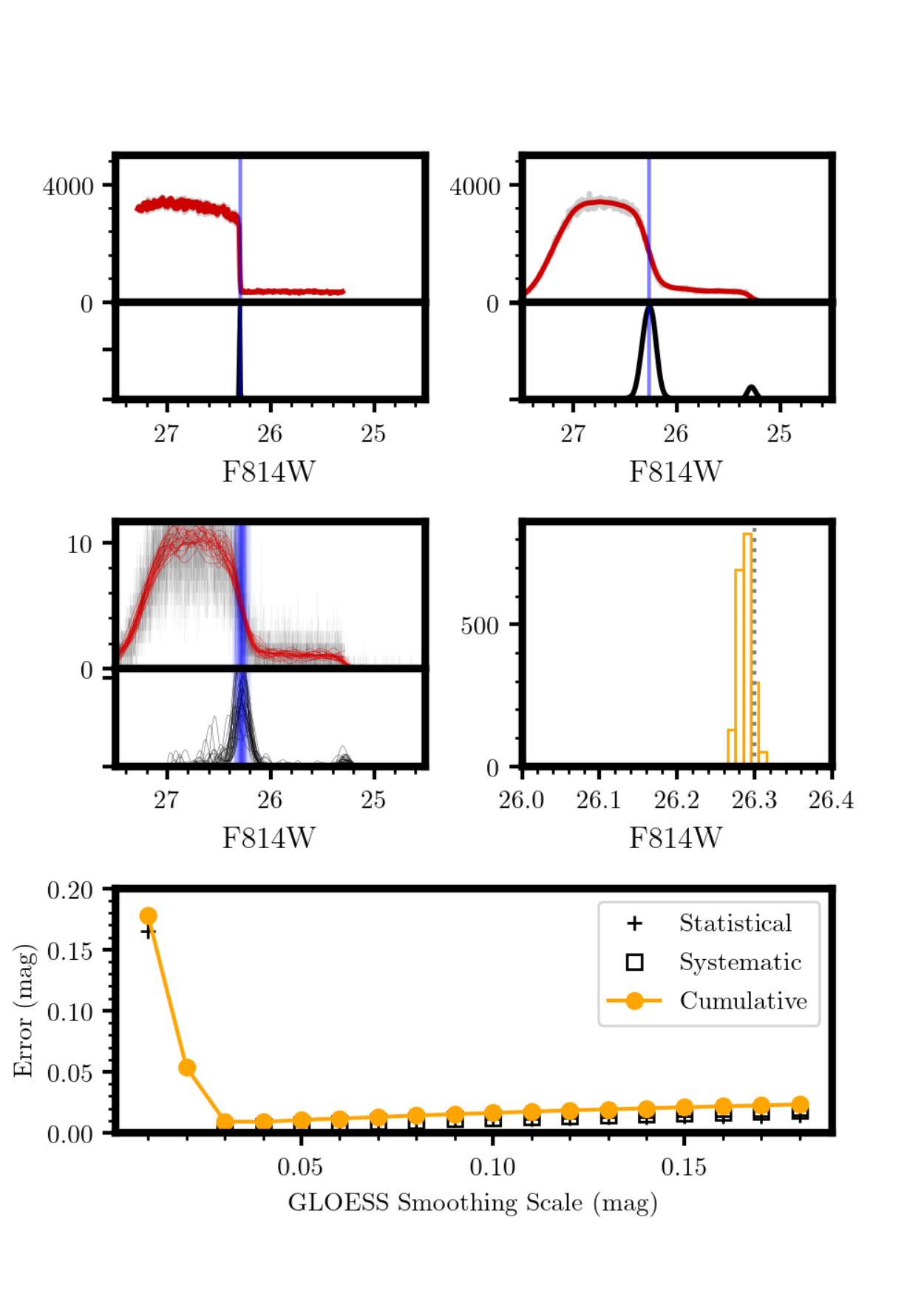}
\caption{Results from the artificial star simulations for M66 and M96. (i) The input theoretical luminosity function and corresponding TRGB response located at exactly the input magnitude, 26.2 and 26.3 mag, for M66 and M96, respectively. (ii) The artificial star luminosity function, constructed by placing artificial stars sampled from the theoretical input from panel (i) into the real data, photometering them, then reconstructing the luminosity function with as-measured magnitudes. Note that incompleteness is not constrained because artificial star positions are fixed. (iii) Overplot of 30 ASLFs downsampled from that shown in panel (ii) to reflect the stellar counts in the observed luminosity function after spatial cuts. The smoothing window size is set to 0.06 mag. (iv) Histogram of individual TRGB detections from 10000 downsampled realizations of the full ASLF depicted in panel (ii), 30 of which are shown in the above panel. The bins have width 0.01 mag. (v) Distribution of statistical and systematic errors as a function of GLOESS smoothing width. Statistical errors are plotted as black plus signs and systematic errors as black squares. The quadrature sum of both, the error in the TRGB measurement, is plotted with red connected circles.
}
\label{fig:art}
\end{figure*}

Our chosen alternative is to simulate the measurement of the TRGB under conditions that directly reflect those of our observed data in a large number of realizations. As such we use the artificial LFs described in \autoref{subsect:artstars} (see panels (i) and (ii) in \autoref{fig:art}), downsampling them to reflect the stellar counts seen in our observed LFs (see panels (iii) in \autoref{fig:art}). We then iteratively ran this empirical realization of our theoretical LF through our TRGB detection algorithm, tabulating the detected TRGB magnitudes for each realization. 

Once a sufficient number of simulated TRGB detections was reached, we could measure the dispersion of this sample of simulated TRGB detections, including any shift of the mean detected magnitude from the input TRGB magnitude. The scatter of this distribution is taken to be the statistical error in our TRGB measurement. The displacement of the mean from the input TRGB magnitude is adopted as the systematic error. Panels (iv) of \autoref{fig:art} display one such distribution of simulated TRGB detections for a single GLOESS smoothing window width. We then repeated this process for a range of smoothing scales, determining the measurement errors as a function of the smoothing window width, the results of which are shown in the bottom panel of \autoref{fig:art}. We adopt for our TRGB detection error budget the point at which the cumulative error vs. smoothing width curve reaches a minimum.

Unlike some of the other targets in the CCHP sample, there is a non-zero systematic magnitude effect in the artificial star photometry that is brought to light by the TRGB simulations for our galaxies. The origin of this offset, which deviates from zero at bright magnitudes and increases to $0.02$~mag at the magnitude range of the TRGB, is unknown. We therefore adopt it as a systematic uncertainty, rather than a bias for which we might be tempted to apply a correction.

\subsection{Contamination by Younger Stellar Populations} \label{subsect:spatialcuts}
In the regions closest to the galaxy center in each of the M66 and M96 pointings, there is a steep gradient in star-like sources. This sharp gradient in stellar density suggests the existence of a younger stellar population(s) in these regions. As a result we detect multiple significant edges in the LF, which itself is not easily separable into distinct AGB, TRGB, and RGB components. In the top row of \autoref{fig:detection} we present the TRGB detections of the full source catalogs, which contain multiple, equal-power edges near the TRGB.

To rectify this contamination by intermediate-age stellar populations we spatially exclude sources located within the red, dashed boundaries indicated in \autoref{fig:pointing}. To determine where to place the spatial cut boundaries we incrementally shifted outward the boundary along lines of constant radius until the detected TRGB magnitude had converged, and that a single significant edge was detected. The TRGB detections resulting from the spatially masked catalogs are shown in the bottom row of \autoref{fig:detection}. The CMDs, accompanying LFs, and TRGB detections are now well-behaved, with a single significant TRGB edge detected.

Finally, we performed a straightforward experiment to further validate the spatial cuts, by inverting the spatial-dependent mask to instead select for the predominantly intermediate-age stars. We confirmed that this increased the contribution from younger stellar populations, introducing multiple significant edges both above and below the magnitude of the single-peaked, halo TRGB response function. As such, we conclude that the spatial-dependent cuts have greatly improved the accuracy of our TRGB determination. 
\subsection{The Distances to M66 and M96} \label{subsect:trgbdist}

In \autoref{fig:detection} we show the TRGB detections for both M66 and M96. The leftmost sub-panels are the CMDs, the middle panels the corresponding marginalized $I$-band LFs, and the rightmost sub-panels contain the signal-to-noise-weighted Sobel response functions, the peak of which is the measured magnitude of the $I$-band TRGB. In the top row of \autoref{fig:detection} we show the TRGB detections before applying the spatial cut discussed in \autoref{subsect:spatialcuts}. In the bottom row are the TRGB detections after applying the spatial cut.

For M66 we measure \( I_{TRGB} =\)\spliterr{\mTRGBMsixtysixRounded}{\mTRGBMsixtysixSTATerrRounded ~(stat)} {\mTRGBMsixtysixSYSerrRounded~(sys)} mag and for M96, \( I_{TRGB} =\)\spliterr{ \mTRGBMninetysixRounded}{\mTRGBMninetysixSTATerrRounded ~(stat)} {\mTRGBMninetysixSYSerrRounded~(sys)}mag. The errors on our TRGB measurements are determined from our artificial star simulations, as already discussed in \autoref{subsect:errors}. The apparent TRGB magnitudes and their associated uncertainties are provided in columns 2 to 4 in \autoref{tab:dists}.

To compute the extinction-corrected TRGB magnitudes for M66 and M96, we assume negligible host reddening in the stellar halos and adopt the \cite{sch11} values for galactic reddening in addition to a \cite{car89} galactic extinction law. The adopted \(I\)-band extinction values are provided in column 5 of \autoref{tab:dists}. Uncertainties on the adopted galactic extinction values are negligible compared to the TRGB measurement uncertainties. We then adopt a TRGB absolute calibration \(M_I^{\mathrm{TRGB}}=\)\spliterr{\trgblum} {\trgblumstaterr ~(stat)} {\trgblumsyserr ~(sys)}~mag., as determined in Freedman et al. (2019, in press). From this, we determine true distance moduli \(\mu_0 =\)\spliterr{\MsixtysixDMRounded}{\MsixtysixCMBstaterrRounded (stat)}{\MsixtysixCMBsyserrRounded (sys)}~mag and \( \mu_0 = \)\spliterr{\MninetysixDMRounded}{\MninetysixCMBstaterrRounded (stat)}{\MninetysixCMBsyserrRounded (sys)}~mag, for M66 and M96, respectively. The final distance moduli and associated uncertainties are provided in columns 6 through 8 of \autoref{tab:dists}. The same values in physical units are provided in columns 9 through 11 of the same table.

\section{Previous Distances to M66 and M96}\label{sect:dists}
 The distances we have measured to these two Leo I group members lie somewhat fainter than previously published distance measurements. These previously published distances are tabulated in \autoref{tab:m66dists} and \autoref{tab:m96dists}. These same distance measurements are displayed in \autoref{fig:dists}. For any Cepheid distance determinations that do not quote a systematic error the global Cepheid PL systematic error of 0.20 mag calculated by the HST Key Project \citep{mad99} should be assumed.

\subsection{M66/NGC~3627} \label{subsect:m66}
\cite{sah99} determined the first Cepheid distance to M66 as part of the Sandage/Saha/Tammann project to calibrate the absolute magnitude of the SNe Ia using HST. Saha et al. used the \cite{mad91} LMC-based Leavitt Law calibration to determine a distance modulus \(\mu_0 = 30.17 \pm 0.12 \) (stat only) mag,  shifting to 30.22 mag after applying the \citet{cas98} long-to-short exposure correction.\footnote{ \cite{dol00} determined that this ``correction'' was due to inconsistent methodologies for calculating the background sky value, but we include it here for historical accuracy.} 

As part of the HST Key Project, \cite{gib00} reanalyzed the \cite{sah99} dataset and found \(\mu_0 = \) \spliterr{30.06}{0.17 (stat)} {0.16 (sys)} using the same \cite{mad91} Leavitt Law calibration. \cite{fre01} then used the \cite{uda99} LMC Leavitt Law calibration to determine \(\mu_0 = \) \spliterr{29.86}{0.08 (stat)} {0.20 (sys)} mag, which increased to \(\mu_0 = \) \spliterr{30.01}{0.15 (stat)} {0.1 (sys)} after applying a \(-0.2\)~mag/dex metallicity correction.

\cite{gib01} re-analyzed the Key Project data with an updated photometric zero point, finding \(\mu_0 = \)\spliterr{29.85}{0.17 (stat)} {0.19 (sys)} mag, and  \(\mu_0 = \)\spliterr{29.99}{0.15 (stat)} {0.1 (sys)} mag, after applying a \(-0.2\) mag/dex metallicity correction. \cite{wil01} found \(\mu_0 = \)\spliterr{29.70}{0.07 (stat)}{0.20 (sys)} and \cite{dol02} determined \(\mu_0 = 30.09~\pm 0.10 \) (stat only)~mag using the \cite{mad91} Leavitt Law and, using the \cite{uda99} Leavitt Law calibration, determined \(\mu_0 = 29.97~\pm 0.09 \) (stat only) mag. \cite{pat02} determined \(\mu_0 = \) 29.80 mag using a Period-Radius relation and adopted a V-band slope to the Leavitt Law as determined from an ensemble of measurements in the LMC. They also determined \(\mu_0 = \) 29.77 mag using a Hipparcos parallax zero point.

\cite{kan03} were the first, in the context of M66, to apply Leavitt Law calibrations not tied to the LMC. They used a total of seven different Leavitt Law calibrations, four based in the LMC \citep{mad91, uda99, tam02, kan03} and three based in the Galaxy \citep{gie98, fou03, tam03}. They found consistently fainter distances when adopting the galactic calibrations, suggesting a metallicity effect, induced by the different calibration samples. For this reason they adopted a -0.2 mag/dex correction to their preferred PL relations, \citep[Milky Way, LMC, respectively]{fou03, kan03}, determining \(\mu_0 = 30.28~\pm 0.11 \) (stat only). Similarly, \citet{sah06} explored the effects of metallicity and period-bias in Cepheid PL relation distances. They determine \(\mu_0 = 30.15\) (stat only) mag and \(\mu_0 = 30.41\)  mag with LMC and galactic Leavitt Law calibrations, respectively. Finally they applied a period-dependent metallicity correction to determine \(\mu_0 = 30.46~\pm 0.09 \) (stat only) mag.

M66 is the second most metal-rich galaxy in the HST KP sample. Its high mean metallicity was thought to systematically offset the distances determined via an LMC-calibrated Leavitt Law, making them \(\sim 0.10 - 0.15 \) mag fainter/closer than those determined using galactic, or metallicity-corrected LMC-based Leavitt Law calibrations.

There have been three prior sets of TRGB distance measurements to M66. \citet{mou09m66} used an archival dataset that was obtained to measure star formation in the disk of M66. It is likely that their edge detector triggered off of the AGB at \spliterr{29.76}{0.10 (stat)}{0.02 (sys)} mag. See \citet{fre89} for an illustrative and comprehensive breakdown of the ambiguous, but well observed, nature of these AGB mis-triggers. \cite{lee13} used an archival HST pointing that targeted Cepheids in the outer disk of M66. After spatially excising the more-crowded, disk-like regions of their fields, they measured \(\mu_0 = \) \spliterr{30.15}{0.03 (stat)}{0.12 (sys)} mag. Following this, \cite{jan17} re-reduced the same dataset using two different geometric anchors, determining \(\mu_0 = \) \spliterr{30.17}{0.04 (stat)}{0.1 (sys)} and \(\mu_0 = \) \spliterr{30.19}{0.04 (stat)}{0.07 (sys) mag}, for the LMC detached eclipsing binary anchor and the NGC 4258 water maser anchor, respectively. \cite{jac09}, as part of the Cosmic Flows Program \citep{tul09,tul13,tul16}, measure for M66 \(\mu_0 =\)\spliterr{30.27}{0.03~(stat)}{0.08~(sys)} mag when using the same dataset as used here.

\autoref{fig:dists} displays the above distance measurements to M66 as a function of publication date. Only Cepheid distances that use metallicity-corrected LMC Leavitt Law calibrations or a galactic calibration are retained in the calculation of the weighted average. The \citet{mou09m66} distance measurement is excluded for being a misidentified TRGB at the bright cutoff of the AGB. The retained distance measurements are shown as filled circles while the excluded distances are shown as open circles. The weighted average of the filled circles is plotted as a gray bar. Note that the filled circles trend significantly fainter/more distant than the open circles. This is a direct result of the metallicity corrections applied to optical \(VI\) Cepheid Leavitt Laws that were calibrated in the LMC.

\subsection{M96/NGC 3368} \label{subsect:m96}
All modern (post-1995) Cepheid distance measurements to M96 make use of the dataset acquired by \cite{tan95}, who observed 13 epochs in the $V$-band and 3 in the $I$-band. They used the \cite{mad91} Leavitt Law calibration to determine \(\mu_0 = \) \(30.32 \pm 0.16\) (stat only) mag. shifting to 30.37 mag after applying the \citet{cas98} long-to-short exposure correction. Later \cite{tan99} added a single epoch to each of their $V$- and $I$-band datasets and used the Leavitt Law calibrations from \cite{tan97}, redetermining the distance modulus to M96 to be \(\mu_0 = \) \(30.13 \pm 0.07\) (stat only). In order to compare directly to their earlier 1995 distance determination, they used the \cite{mad91} PL relations and found internal agreement with \(\mu_0 = \) \(30.16 \pm 0.07\) (stat only) mag. Finally they applied a Cepheid zero-point metallicity correction of -0.3~mag/dex, increasing their distance measurement by 0.12 mag to \(\mu_0 = 30.25 \pm 0.18 \) (total error) mag. 

The HST Key Project began the next focused study of M96. First \cite{gib00} used the \cite{mad91} LMC-based Leavitt Laws to determine \(\mu_0 = \)\spliterr{30.20}{0.10 (stat)}{0.16 (sys)} mag. Next \cite{kel00} applied the \cite{ken98} -0.24 mag/dex zero-point metallicity correction, giving \(\mu_0 = 30.37\) mag. Finally, \cite{fre01} adopted the OGLE Leavitt Laws \citep{uda99} and included a period cut to avoid a period-dependent bias. As a result they determined \(\mu_0 = \) \(29.97 \pm 0.06\) (stat only). They then applied a -0.2 mag/dex metallicity correction for a final distance modulus of \(\mu_0 = \) \(30.11 \pm 0.2\) mag (cumulative error).

The last of the Cepheid distance measurements to M96 are a pair of re-analyses that used a combination of the above two datasets. \cite{pat02} first used the Baade-Wesselink method to calibrate the Leavitt Law zero-point and measured \(\mu_0 = \) \(30.17 \pm 0.10 \) (stat only) mag. Using a Hipparcos calibration of the Leavitt Law, they determined \(\mu_0 = \) \(30.05 \pm 0.06\) (stat only) mag. Similarly, \cite{sah06} reanalyzed both the \cite{tan95} and Key Project datasets in order to explore the effects of various zero-point calibrations on their final distance measurement. Firstly, via an LMC-based Leavitt Law calibration they determined a distance \(\mu_0 = 29.98\) mag. They then used a galactic Leavitt Law calibration and measured \( \mu_0 = 30.25 \) mag. Finally, from comparing the galactic and LMC PL relations they applied a period-dependent metallicity correction and determined a final \( \mu_0 = 30.30 \pm 0.11 \) (stat only) mag. 

There have been three previous TRGB distance measurements to M96. \cite{mou09m96} determine \(29.65 \pm 0.28 \) mag. This appears to be a disconcertingly bright result until one considers the positioning of the archival field used: it is squarely in the disk of M96. Upon closer inspection, the TRGB can be seen in the response function presented in their Figure 4 as a broad bump of relatively low amplitude, located much fainter than their adopted TRGB magnitude. The evidence for this result being a mis-trigger off of younger stellar populations is twofold: Firstly, the authors elect to use the log of the luminosity function and do not apply a signal-to-noise weight to their edge response function, thus increasing the power in sparsely populated regions of the luminosity function. Secondly, because the data were taken in the disk, there is an unavoidably significant amount of contamination by intermediate-aged stars (which are apparent in their CMD and the accompanying I-band luminosity function, their Figure 4). For these reasons the \cite{mou09m96} result is omitted in our cumulative distance analysis and serves as a cautionary tale against application of the TRGB method to regions of recent star formation.

Most recently \cite{lee13} measure \(\mu_0 = \)\spliterr{30.15}{0.03}{0.12} mag. \cite{jan17} then re-calibrate the same dataset, with improvements made to aperture corrections as well as an additional metallicity correction, to measure \(\mu_0 = \)\spliterr{30.17}{0.04}{0.06} mag, a result that agrees with ours. As a check we re-ran this completely independent dataset through our own pipeline and confirmed the agreement between the two distance measurements.

Finally, the Cosmicflows-3 program \citep{tul16}, using the methodology of \cite{jac09}, measured \(\mu_0 =\) \spliterr{30.25}{0.02~(stat)}{0.08~(sys)} mag, again using the same HST data used in the present paper. This agreement is reassuring, especially when considering the differences in methodology: here we use DAOPHOT photometry and a Sobel edge detector, while the EDD employs DOLPHOT and a maximum likelihood fit to the TRGB. Ignoring differences in the adopted absolute zero point of the TRGB, our results agree at the 1\% level.

\begin{figure*}
\includegraphics[width = 0.49\textwidth]{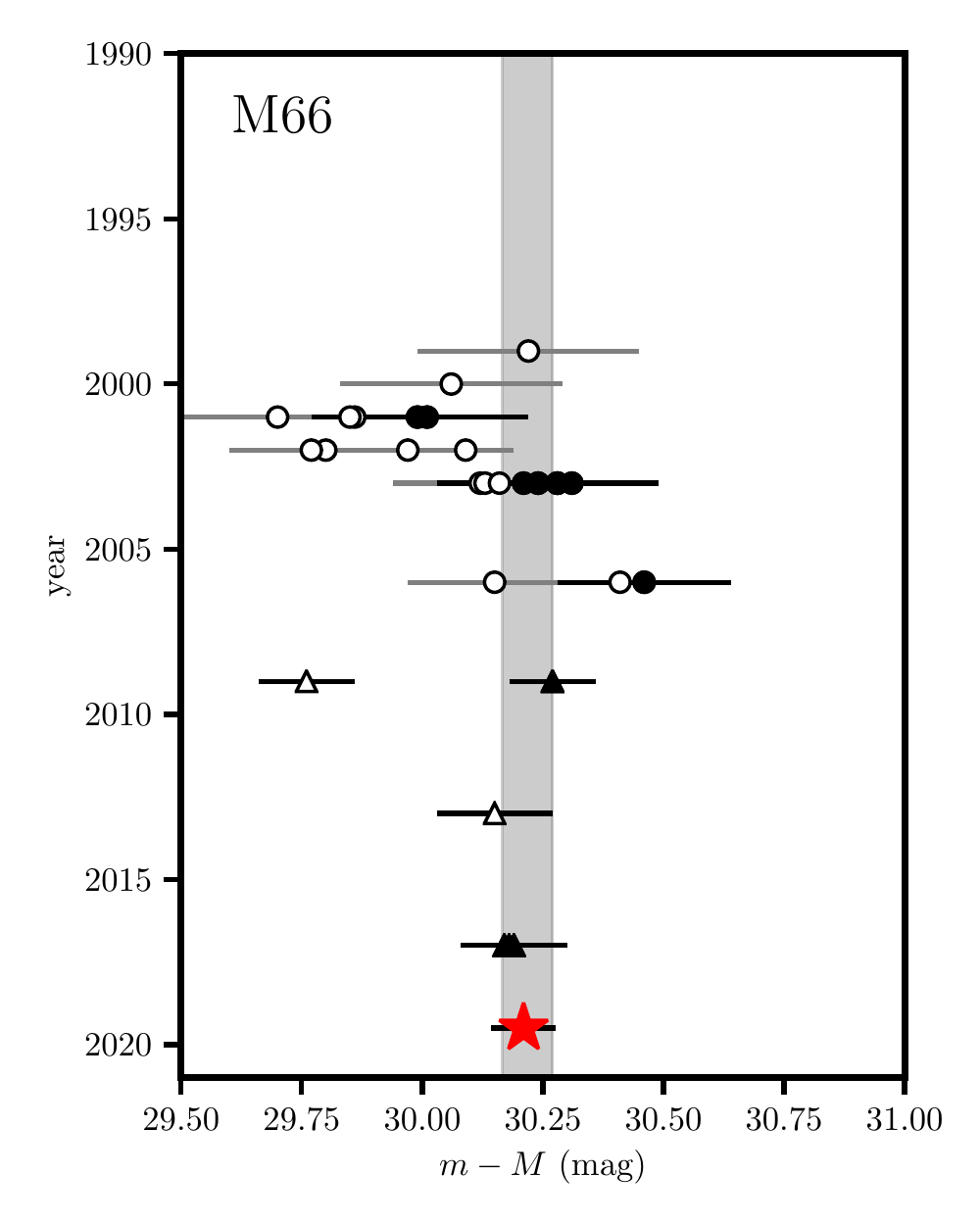}
\includegraphics[width = 0.49\textwidth]{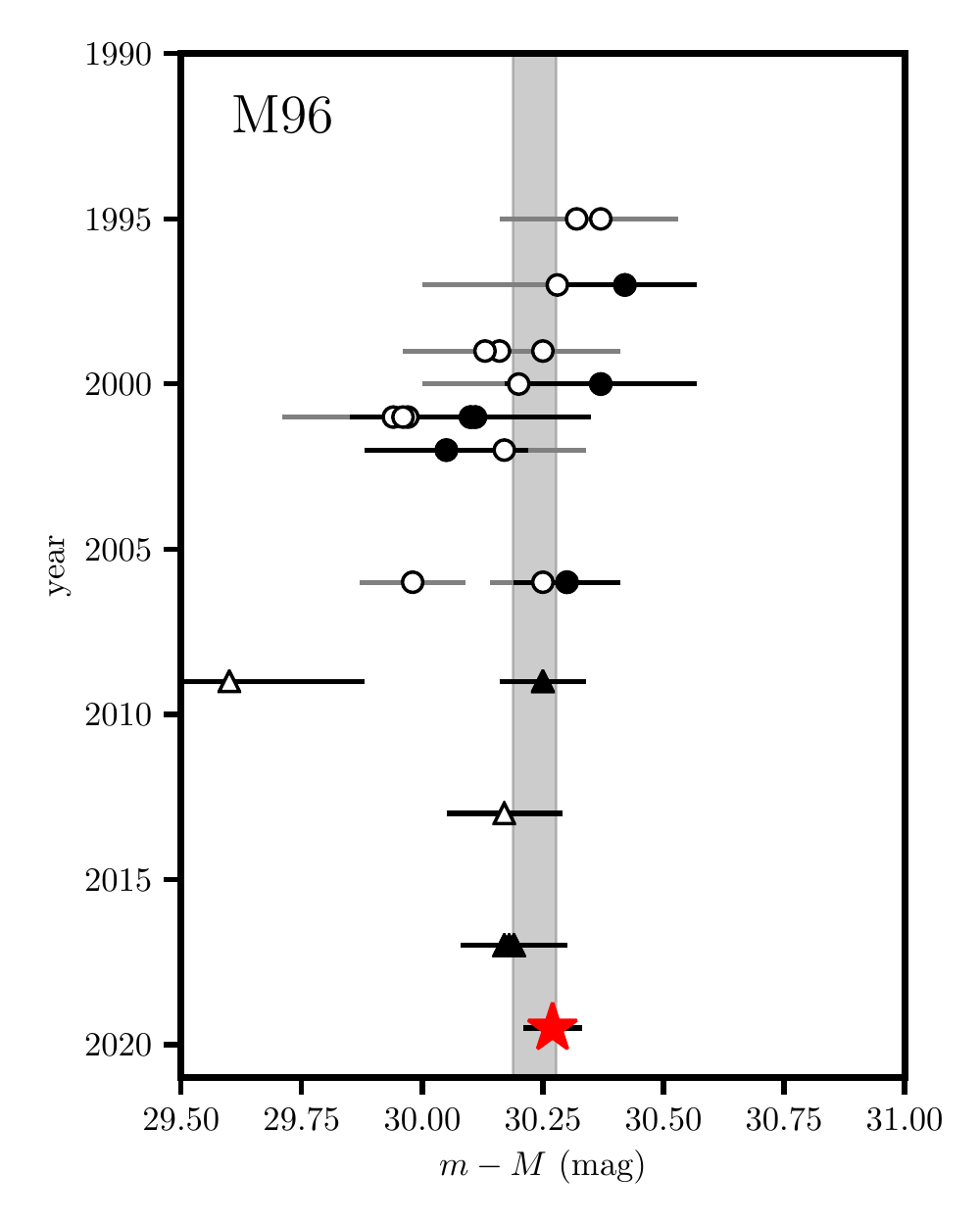}
\caption{\textit{left}: Literature Cepheid and TRGB distance measurements to M66, beginning in 1990. \textit{right}: The same for M96. Circles represent distances determined using Cepheids, triangles using the TRGB. The red star in each figure is the distance modulus presented here. Filled markers represent the values used in calculating the weighted mean distance, plotted as a gray band of width equal to the error on the mean. Open markers are distance moduli with less favored assumptions for, e.g. the zero-point or metallicity correction term.}
\label{fig:dists}
\end{figure*}

The apparent wide range (\(\sim0.5 \) mag) of Cepheid distance determinations is ostensibly due to large metallicity corrections for M96, as was also the case for M66. When considering only the metal-corrected or galactic-calibrated Cepheid distance measurements, there is an apparent convergence centered around \( \mu_0 = 30.25 \) mag, as evidenced in \autoref{fig:dists}.

\input{./cchp_distance_table.tex}

\input{./M66_literature_distance_table.tex}

\input{./M96_literature_distance_table.tex}

\section{Summary and Conclusions} \label{sect:conc}
With the TRGB we have measured the distances to a pair of Leo I Group galaxies, M66 and M96. For M66 we determine \(\mu_0 = \) \spliterr{\MsixtysixDMRounded}{\MsixtysixCMBstaterrRounded~(stat)}{\MsixtysixCMBsyserrRounded~(sys)}~mag and for M96 \( \mu_0 = \)\spliterr{\MninetysixDMRounded}{\MninetysixCMBstaterrRounded~(stat)}{\MninetysixCMBsyserrRounded~(sys)}~mag.
Our measured distances agree well with both published metallicity-corrected Cepheid distances as well as properly measured TRGB distances.

In light of conflicting determinations of systematic effects of metallicity on the multi-wavelength Leavitt Law zero-points (see, e.g. \cite{sto11, gie18} for studies that measure non-zero metallicity effects and \cite{per04} for systematic effects in agreement with zero), the importance of high-precision, stellar-halo-based, Population~II distance measurements to M66 and M96, like those presented here, cannot be overstated. This is underlined in \autoref{fig:dists}, in which the error on the mean of the distribution of metallicity-corrected Cepheid distances, plotted in gray, is comparable to the error in our \emph{single} measurement using the TRGB, in red, while the dispersion of the Cepheid distance distribution is an order of magnitude larger, dominated by the wide range of metallicity corrections adopted, if any. This suggests that, in environments with metal abundances that deviate significantly from galactic or LMC metallicities, a single distance measurement using the stellar-halo-based TRGB could be a factor of several more efficient and accurate than a single Cepheid-based distance measurement.

The high extinction and high mean metallicity in the disks of the Leo Group galaxies M66 and M96 has henceforth precluded their inclusion in modern, post-HST Key Project, measurements of the Hubble Constant. However, through improved multi-wavelength SN Ia light curve fitting techniques, combined with the high-precision Population II distances determined here, the pair becomes viable. This particular case emphasizes the power of a proper application of the TRGB in the context of the extragalactic distance scale. To minimize its own potential systematics, e.g. mixed populations of mass, age, and metallicity and their resulting effects on the TRGB luminosity (if not calibrated for), the Population II TRGB must be measured in the stellar halo (see Beaton et al. 2019, submitted, for an extensive discussion on this topic in the context of M101). In particularly metal-rich or particularly metal-poor environments, an author may or may not make a still-debated metallicity correction to a Cepheid distance measurement, leading to a large dispersion in measured distances. By contrast, measuring the TRGB in the stellar halo of galaxies avoids this problem altogether, allowing for more accurate distance measurements.

We have shown here that, through careful application, the TRGB method to determine extragalactic distances has the potential to provide distances accurate to a few percent out to about 20 Mpc, pending a future sub-percent determination of the TRGB absolute magnitude using GAIA parallaxes. The halo-based requirement of extragalactic applications of the TRGB, coupled with the fact that the TRGB luminosity is a factor of 6 brighter in the NIR, will allow for similarly precise distances out to yet further distances, likely pushing the TRGB's reach to 50 Mpc, and hopefully further with James Webb. The implications are immediately obvious for the extragalactic distance scale and, hence, accurate, 1\% level measurements of the Hubble Constant.

The authors greatly appreciate and thank Mark Seibert for seminal contributions to this manuscript and the CCHP. We thank Peter Stetson for a copy of DAOPHOT. M.G.L. was supported by the National Research Foundation of Korea (NRF) grant funded by the Korea Government (MSIP) No. 2017R1A2B4004632. Support for this work was provided by NASA through Hubble Fellowship grant \#51386.01 awarded to R.L.B. by the Space Telescope Science Institute, which is operated by the Association of  Universities for Research in Astronomy, Inc., for NASA, under contract NAS 5-26555. Support for program \#13691 was provided by NASA through a grant from the Space Telescope Science Institute, which is operated by the Association of Universities for Research in Astronomy, Inc., under NASA contract NAS 5-26555. Some of the data presented in this paper were obtained from the Mikulski Archive for Space Telescopes (MAST). STScI is operated by the Association of Universities for Research in Astronomy, Inc., under NASA contract NAS5-26555. This research has made use of the NASA/IPAC Extragalactic Database (NED) which is operated by the Jet Propulsion Laboratory, California Institute of Technology, under contract with the National Aeronautics and Space Administration. This research has made use of the NASA/IPAC Infrared Science Archive (IRSA), which is operated by the Jet Propulsion Laboratory, California Institute of Technology, under contract with the National Aeronautics and Space Administration. 
Facility: HST (ACS/WFC).
Software: DAOPHOT \citep{ste87}, ALLFRAME \citep{ste94}, TinyTim \citep{kri11}.

\bibliographystyle{aasjournal}
\bibliography{main.bbl}

\end{document}

%% file: measurements.tex
\newcommand{\mTRGBMsixtysix}{26.23}
\newcommand{\mTRGBMninetysix}{26.28}

\newcommand{\mTRGBMsixtysixSTATerr}{0.03}
\newcommand{\mTRGBMninetysixSTATerr}{0.01}


%% file: calibration_constants.tex
\newcommand{\trgblum}{-4.049}
\newcommand{\trgblumstaterr}{0.022} 
\newcommand{\trgblumsyserr}{0.039}

\newcommand{\VextinctionMsixtysix}{0.087} 
\newcommand{\VextinctionMninetysix}{0.069} 
\newcommand{\IextinctionMsixtysix}{0.049} 
\newcommand{\IextinctionMninetysix}{0.039}

\newcommand{\ZPerr}{0.02}
\newcommand{\EEerr}{0.02}
\newcommand{\Apcorrerr}{0.03}

%% file: calculations.tex
\FPeval{\CalibrationErr}{ (\ZPerr^2+\EEerr^2+\Apcorrerr^2)^0.5 }

\FPeval{\mTRGBMsixtysixSYSerr}{ (0.02^2+\CalibrationErr^2)^0.5}
\FPeval{\mTRGBMninetysixSYSerr}{ (0.01^2+\CalibrationErr^2)^0.5}

\FPeval{\MsixtysixDM}{\mTRGBMsixtysix-\IextinctionMsixtysix-\trgblum}
\FPeval{\MninetysixDM}{\mTRGBMninetysix-\IextinctionMninetysix-\trgblum}

\FPeval{\MsixtysixDist}{ 10^( (\MsixtysixDM) /5)/100000 }
\FPeval{\MninetysixDist}{ 10^( (\MninetysixDM) /5)/100000 } 

\FPeval{\MsixtysixmTRGBerrCMB}{ (\mTRGBMsixtysixSTATerr^2+\mTRGBMsixtysixSYSerr^2+\CalibrationErr^2)^0.5 }
\FPeval{\MninetysixmTRGBerrCMB}{ (\mTRGBMninetysixSTATerr^2+\mTRGBMninetysixSYSerr^2+\CalibrationErr^2)^0.5 }

\FPeval{\MsixtysixCMBstaterr}{(\trgblumstaterr^2+\mTRGBMsixtysixSTATerr^2)^0.5  }
\FPeval{\MsixtysixCMBsyserr}{(\trgblumsyserr^2+\mTRGBMsixtysixSYSerr^2+(\IextinctionMsixtysix/2)^2)^0.5  }
\FPeval{\MsixtysixCMBerr}{ (\MsixtysixCMBstaterr^2+\MsixtysixCMBsyserr^2)^0.5 }

\FPeval{\MninetysixCMBstaterr}{(\trgblumstaterr^2+\mTRGBMninetysixSTATerr^2)^0.5  }
\FPeval{\MninetysixCMBsyserr}{(\trgblumsyserr^2+\mTRGBMninetysixSYSerr^2+(\IextinctionMninetysix/2)^2)^0.5  }
\FPeval{\MninetysixCMBerr}{ (\MninetysixCMBstaterr^2+\MninetysixCMBsyserr^2)^0.5 }

\FPeval\MsixtysixDistupperrdiststat{ 10^( (\MsixtysixDM+\MsixtysixCMBstaterr) /5)/100000 }
\FPeval\MsixtysixDistlowerdiststat{ 10^( (\MsixtysixDM-\MsixtysixCMBstaterr) /5)/100000 }
\FPeval\MsixtysixDiststaterr{ 0.5*(\MsixtysixDistupperrdiststat - \MsixtysixDistlowerdiststat) }
\FPeval\MninetysixDistupperrdiststat{ 10^( (\MninetysixDM+\MninetysixCMBstaterr) /5)/100000 }
\FPeval\MninetysixDistlowerdiststat{ 10^( (\MninetysixDM-\MninetysixCMBstaterr) /5)/100000 }
\FPeval\MninetysixDiststaterr{ 0.5*(\MninetysixDistupperrdiststat - \MninetysixDistlowerdiststat) }
\FPeval\MsixtysixDistupperrdistsys{ 10^( (\MsixtysixDM+\MsixtysixCMBsyserr) /5)/100000 }
\FPeval\MsixtysixDistlowerdistsys{ 10^( (\MsixtysixDM-\MsixtysixCMBsyserr) /5)/100000 }
\FPeval\MsixtysixDistsyserr{ 0.5*(\MsixtysixDistupperrdistsys - \MsixtysixDistlowerdistsys) }
\FPeval\MninetysixDistupperrdistsys{ 10^( (\MninetysixDM+\MninetysixCMBsyserr) /5)/100000 }
\FPeval\MninetysixDistlowerdistsys{ 10^( (\MninetysixDM-\MninetysixCMBsyserr) /5)/100000 }
\FPeval\MninetysixDistsyserr{ 0.5*(\MninetysixDistupperrdistsys - \MninetysixDistlowerdistsys) }

\FPeval{\distanceModDiff}{abs(\MsixtysixDM-\MninetysixDM)}
\FPeval{\distanceModDifferr}{ (\mTRGBMsixtysixSTATerr^2+\mTRGBMninetysixSTATerr^2+\mTRGBMsixtysixSYSerr^2+\mTRGBMninetysixSYSerr^2 + (\IextinctionMsixtysix/2)^2 + (\IextinctionMninetysix/2)^2+ 2*\Apcorrerr^2)^0.5}

\FPeval{\distanceDiff}{ abs(\MninetysixDist - \MsixtysixDist) }
\FPeval{\distanceDifferr}{ (\MninetysixDiststaterr^2 + \MsixtysixDiststaterr^2)^0.5 }

%% file: document_variables.tex
\FPeval{\mTRGBMsixtysixRounded}{round(\mTRGBMsixtysix,2)}
\FPeval{\mTRGBMninetysixRounded}{round(\mTRGBMninetysix,2)}

\FPeval{\mTRGBMsixtysixSTATerrRounded}{round(\mTRGBMsixtysixSTATerr,2)}
\FPeval{\mTRGBMninetysixSTATerrRounded}{round(\mTRGBMninetysixSTATerr,2)}
\FPeval{\mTRGBMsixtysixSYSerrRounded}{round(\mTRGBMsixtysixSYSerr,2)}
\FPeval{\mTRGBMninetysixSYSerrRounded}{round(\mTRGBMninetysixSYSerr,2)}

\FPeval{\MsixtysixCMBstaterrRounded}{round(\MsixtysixCMBstaterr,2)}
\FPeval{\MsixtysixCMBsyserrRounded}{round(\MsixtysixCMBsyserr,2)}
\FPeval{\MsixtysixCMBerrRounded}{round(\MsixtysixCMBerr,2)}
\FPeval{\MninetysixCMBstaterrRounded}{round(\MninetysixCMBstaterr,2)}
\FPeval{\MninetysixCMBsyserrRounded}{round(\MninetysixCMBsyserr,2)}
\FPeval{\MninetysixCMBerrRounded}{round(\MninetysixCMBerr,2)}

\FPeval{\VextinctionMsixtysixRounded}{round(\VextinctionMsixtysix,2)} 
\FPeval{\VextinctionMninetysixRounded}{round(\VextinctionMninetysix,2)} 
\FPeval{\IextinctionMsixtysixRounded}{round(\IextinctionMsixtysix,2)} 
\FPeval{\IextinctionMninetysixRounded}{round(\IextinctionMninetysix,2)}

\FPeval{\MsixtysixDMRounded}{round(\MsixtysixDM,2)}
\FPeval{\MninetysixDMRounded}{round(\MninetysixDM,2)}

\FPeval{\MsixtysixDistRounded}{round(\MsixtysixDist,2)}
\FPeval{\MninetysixDistRounded}{round(\MninetysixDist,2)}

\FPeval{\MsixtysixDiststaterrRounded}{round(\MsixtysixDiststaterr,2)}
\FPeval{\MninetysixDiststaterrRounded}{round(\MninetysixDiststaterr,2)}
\FPeval{\MsixtysixDistsyserrRounded}{round(\MsixtysixDistsyserr,2)}
\FPeval{\MninetysixDistsyserrRounded}{round(\MninetysixDistsyserr,2)}

\FPeval{\distanceModDiffRounded}{round(\distanceModDiff,2)}
\FPeval{\distanceModDifferrRounded}{ round(\distanceModDifferr,2) }

\FPeval{\distanceDiffRounded}{round(1000*round(\distanceDiff,2),0)}
\FPeval{\distanceDifferrRounded}{round(1000*round(\distanceDifferr,2),0)}

%% file: obs_log.tex
\begin{deluxetable*}{cccccccc} 
\tabletypesize{\normalsize} 
\tablewidth{0pt} 
\tablecaption{ACS/WFC Observation Summary\label{tab:obs}} 
\tablehead{ 
\colhead{Target} &
\colhead{Dates} &
\colhead{F606W}&
\colhead{F814W} &
\colhead{RA (h:m:s)\textsuperscript{a}} &
\colhead{Dec (d:m:s)\textsuperscript{a}} &
\colhead{Field Size} 
}
\startdata 
M66 (NGC~3627) & 2015-03-10 &  $1 \times 2400$s & $2\times 2400$s &  11:20:31.3 & +12:59:56.3 & $3'37''\times 3'37''  $  \\
M96 (NGC~3368)& 2015-11-29 &  $1\times 2400$s &  $3\times 2400$s &  10:47:05.1 & +11:49:22.6 & $3'37''\times 3'37''  $ \\
\enddata 
\tablecomments{\textsuperscript{a}J2000 coordinates at the center of the HST pointings, as shown in red in \autoref{fig:pointing}.}
\end{deluxetable*} 

%% file: apcors.tex
\begin{deluxetable*}{ccccccccc} 
\tabletypesize{\normalsize} 
\tablewidth{0pt} 
\tablecaption{Average measured aperture corrections at 0.5'' \label{tab:apcors}} 
\tablehead{ 
\colhead{Target} &
\multicolumn{2}{c}{chip 1}  &
\multicolumn{2}{c}{chip 2}   \\
\colhead{} &
\colhead{F606W} & \colhead{F814W} &
\colhead{F606W} & \colhead{F814W}
}
\startdata 
M66 & $-0.04(08)$ & $-0.17(02)$ & $ 0.01(08)$ & $-0.09(04)$ \\
M96 & $-0.11(08)$ & $-0.11(02)$ & $-0.03(05)$ & $-0.07(03)$\\
\enddata 
\tablecomments{The values in parentheses are the standard deviations for each set of frame-by-frame aperture corrections.} 
\end{deluxetable*}

%% file: cchp_distance_table.tex
\begin{deluxetable*}{cccccccccccc} 
\tabletypesize{\normalsize} 
\tablewidth{0pt} 
\tablecaption{TRGB magnitudes, foreground reddenings, and true distances to M66 and M96. \label{tab:dists}} 
\tablehead{ 
\colhead{Galaxy} &
\colhead{$m_{\mathrm{TRGB}}$\tablenotemark{a}} &
\colhead{$\sigma_{stat}$} &
\colhead{$\sigma_{sys}$} &
\colhead{$A_{\mathrm{F814W}}$} &
\colhead{$\left(m-M\right)_0$\tablenotemark{b}} &
\colhead{$\sigma_{stat}$} &
\colhead{$\sigma_{sys}$} &
\colhead{$D$~(Mpc)} &
\colhead{$\sigma_{stat}$} &
\colhead{$\sigma_{sys}$} &
}
\startdata 
M66 & \mTRGBMsixtysixRounded & \mTRGBMsixtysixSTATerrRounded & \mTRGBMsixtysixSYSerrRounded & \IextinctionMsixtysixRounded & \MsixtysixDMRounded & \MsixtysixCMBstaterrRounded & \MsixtysixCMBsyserrRounded & \MsixtysixDistRounded & \MsixtysixDiststaterrRounded & \MsixtysixDistsyserrRounded \\
M96 & \mTRGBMninetysixRounded & \mTRGBMninetysixSTATerrRounded & \mTRGBMninetysixSYSerrRounded & \IextinctionMninetysixRounded & \MninetysixDMRounded & \MninetysixCMBstaterrRounded & \MninetysixCMBsyserrRounded & \MninetysixDistRounded & \MninetysixDiststaterrRounded & \MninetysixDistsyserrRounded \\
\enddata
\tablenotetext{a}{F814W}
\tablenotetext{b}{\(M_{\text{F814W}}^{\mathrm{TRGB}}=\)\spliterr{\trgblum} {\trgblumstaterr ~(stat)} {\trgblumsyserr ~(sys)}~mag.}
\end{deluxetable*} 

%% file: M66_literature_distance_table.tex
\begin{deluxetable*}{ccccc} 
\tabletypesize{\normalsize} 
\tablecaption{Literature Cepheid and TRGB distances to M66.\label{tab:m66dists}} 
\tablehead{
\colhead{Reference} &
\colhead{\( \mu \) (mag) } &
\colhead{\( \sigma \) (stat) } &
\colhead{\( \sigma \) (sys) } &
\colhead{Notes}
}
\startdata 
\hline
Cepheids \\
\hline 
\cite{sah99} & 30.22 & 0.12 & \ldots  & (1) \\
\cite{gib00} & 30.06 & 0.17 & 0.16 &  (1) \\
\cite{fre01} & 29.86 & 0.08 & 0.20 & (2)      \\
             & 30.01 & 0.17 & 0.13 & (2,4)    \\
\cite{gib01} & 29.85 & 0.17 & 0.19 & (2)      \\
		     & 29.99 & 0.15 & 0.1  & (2,4)    \\ 
\cite{wil01} & 29.70 & 0.07 & 0.20 & (2)    \\
\cite{dol02} & 30.09 & 0.10 & \ldots & (1)  \\
             & 29.97 & 0.09 & \ldots & (2) \\
\cite{pat02} & 29.80 & 0.06 & \ldots & (1,5)     \\
             & 29.77 & 0.07 & \ldots & (6) \\
\cite{kan03} & 30.24 & 0.08 & \ldots & (1)    \\
		     & 30.13 & 0.08 & \ldots & (2)    \\
		     & 30.13 & 0.08 & \ldots & (7)    \\
		     & 30.16 & 0.08 & \ldots & (8)    \\
		     & 30.24 & 0.08 & \ldots & (9)    \\
		     & 30.21 & 0.08 & \ldots & (10)   \\
		     & 30.31 & 0.08 & \ldots & (11)   \\
		     & 30.28 & 0.11 & \ldots & (4,7)  \\
		     & 30.28 & 0.11 & \ldots & (4,10) \\
\cite{sah06}   & 30.15 & \ldots & \ldots & (12) \\
		     & 30.41 & \ldots & \ldots & (13) \\
		     & 30.46 & 0.09 & \ldots & (12,13,14) \\
\hline
TRGB \\
\hline
\cite{mou09m66} & 29.76 & 0.10 & 0.02 & (15) \\
\cite{jac09} & 30.27 & 0.03 & 0.08 &  \\
\cite{lee13} & 30.15 & 0.03 & 0.12 &  (16)\\
\cite{jan17} & 30.17 & 0.04 & 0.10 &  (17) \\
             & 30.19 & 0.04 & 0.07 &  (18) \\
             & 30.18 & 0.04 & 0.06 &  (19)\\
\hline
CCHP & \MsixtysixDMRounded  & \MsixtysixCMBstaterrRounded & \MsixtysixCMBsyserrRounded &
\enddata
\tablecomments{All Cepheid distances make use of the \cite{sah99} dataset and are brought onto a common LMC modulus \(\mu_0\) = 18.5. (1) \cite{mad91} Leavitt Law (2) \cite{uda99} Leavitt Law (3) \cite{ken98} \(-0.24\) mag/dex metal correction; (4) \(-0.2\) mag/dex metal correction; (5) Period-radius relation zero point; (6) Hipparcos zero point (7) \cite{kan03} Leavitt Law; (8) \cite{tam02}; (9) \cite{gie98} Leavitt Law; (10) \cite{fou03} Leavitt Law; (11) \cite{tam03} Leavitt Law; (12) \cite{nge05} LMC Leavitt Law and zero point; (13) \cite{nge04} galactic Leavitt Law and zero point; (14) Period-dependent metal correction; (15) Uses data from \citet[HST PID:10402]{prop:cha04}  (16) Uses data from \citet[HST PID:10433]{prop:mad04}  (17) LMC calibration of TRGB absolute magnitude; (18) NGC 4258 calibration of TRGB absolute magnitude; (19) Combined LMC+NGC4258 TRGB calibration. }
\end{deluxetable*}

%% file: M96_literature_distance_table.tex
\begin{deluxetable*}{ccccc} 
\tabletypesize{\normalsize} 
\tablewidth{0pt} 
\tablecaption{Literature Cepheid and TRGB distances to M96. \label{tab:m96dists}} 
\tablehead{
\colhead{Reference} &
\colhead{\( \mu \)  (mag) } &
\colhead{\( \sigma \) (stat) } &
\colhead{\( \sigma \) (sys) } &
\colhead{Notes}
}

\startdata 
\hline
Cepheids \\
\hline 
\cite{tan95} & 30.32 & 0.16 & \ldots & (1) \\
             & 30.37 & 0.16 & \ldots & (1,2) \\
\cite{koc97} & 30.28 & 0.28 & \ldots   &       \\
             & 30.42 & 0.15 & \ldots   &       \\
\cite{tan99} & 30.16 & 0.07 & \ldots & (1,2) \\
             & 30.13 & 0.07 & \ldots & (2,3) \\
             & 30.25 & 0.18 & \ldots & (2,3,4) \\
\cite{gib00} & 30.20 & 0.10 & 0.16   & (1) \\
\cite{kel00} & 30.37 & 0.10 & 0.20  & (1,5) \\
\cite{wil01} & 29.94 & 0.13 & 0.20 & (2)    \\
\cite{fre01} & 29.97 & 0.06 & 0.20   & (6) \\ 
             & 30.11 & 0.20 & 0.14   & (6,7) \\
\cite{gib01} & 30.01 & 0.10 & 0.18   & (6) \\
		     & 30.15 & 0.10 & 0.18   & (6,7) \\ 
\cite{pat02} & 30.17 & 0.1  & \ldots  & (1,8) \\
             & 30.05 & 0.06 & \ldots   & (1,9) \\
\cite{sah06} & 29.98 & \ldots & \ldots & (10) \\
             & 30.25 & \ldots & \ldots & (11) \\
             & 30.30 & 0.11 & \ldots & (10,11,12) \\
\hline
TRGB \\
\hline
\cite{mou09m96} & 29.65 & 0.28 & \ldots & (13) \\
\cite{jac09} & 30.25 & 0.02 & 0.08 &  \\
\cite{lee13} & 30.15 & 0.03 & 0.12 & (14)\\
\cite{jan17} & 30.21 & 0.05 & 0.10 & (15) \\
             & 30.23 & 0.05 & 0.07 & (16) \\
             & 30.22 & 0.05 & 0.06 & (17)\\
\hline
CCHP & \MninetysixDMRounded & \MninetysixCMBstaterrRounded & \MninetysixCMBsyserrRounded & 
\enddata 
\tablecomments{All Cepheid-based distances use the combined \cite{tan95} and \cite{tan99} dataset and are brought onto a common LMC modulus \(\mu_0\) = 18.5. (1) \cite{mad91} Leavitt Law (2) 0.05 mag ``short-to-long exposure" correction \citep{cas98}; (3) \cite{tan99b} Leavitt Law; (4) \(-0.30\) mag/dex metal correction; (5) \cite{ken98} \(-0.24\) mag/dex metal correction; (6) \cite{uda99} Leavitt Law; (7) \(-0.2\) mag/dex metal correction; (8) Baade-Wesselink Period-radius relation zero point; (9) Hipparcos zero point; (10) \cite{nge05} LMC Leavitt Law and zero point; (11) \cite{nge04} galactic Leavitt Law and zero point; (12) Period-dependent metal correction; (13) Uses data from \citet[HST PID:5415] {prop:tan94} (14) Uses data from \citet[HST PID:10433]{prop:mad04} (15) LMC calibration of TRGB absolute magnitude; (16) NGC~4258 calibration of TRGB absolute magnitude; (17) Combined LMC+NGC~4258 TRGB calibration. }
\end{deluxetable*} 